# Limits of responsiveness concerning human-readable knowledge bases: an operational analysis

G.C. Pentzaropoulos

*Mathematics and Information Technology Unit, Department of Economics, University of Athens, 8 Pesmazoglou Street, 105 59 Athens, Greece.*

______________________________________________________________________

## *Abstract*

***Introduction.*** The purpose of this work is the evaluation of responsiveness when remote users communicate with a human-readable knowledge base (KB). Responsiveness [$R(s)$] is considered here as a measure of service quality.
***Method.*** The preferred method is operational analysis, a variation of classical stochastic theory, which allows for the study of user-system interaction with minimal computational effort.
***Analysis.*** The analysis is based on well-known performance metrics, such as service ability, elapsed time, and throughput: from these metrics estimates of $R(s)$ are derived analytically.
***Results.*** Critical points indicating congestion are obtained: these are limits on the number of admissible requests and the number of connected users. Also obtained is a sufficient condition for achieving flow balance between the KB host and the request-relaying servers.
***Conclusions.*** When $R(s)$ is within normal limits, users should appreciate the benefits from using the services offered by their KB host. When bottlenecks are formed, $R(s)$ declines, and the whole communication system heads for saturation. Flow balancing procedures are necessary for the elimination of bottlenecks, which leads to a better resource management.

## Introduction

Knowledge bases are essential elements of the emerging knowledge economy and society. The efficiency of today's knowledge bases (KBs) relies largely on the capabilities of high-speed telecommunications networks. Available features include broadband infrastructures, intelligent access forms, and improved quality of service. Both state authorities and the Internet providers now offer a wide range of services in a full competion environment. Such services include on-line searching in digital libraries, the transfer of files and video, as well as more advanced forms of collaborative work, such as multicast conferencing and live decision-making across transnational borders. Developments have been strongly supported by significant regulatory and standardization work especially in the European Union and in other parts of the industrial world (European Commission 2002).

KBs are often classified as either machine-readable or human-readable bases. The former contain rules used for automated reasoning under certain conditions. The latter, which are examined here, are used for retrieval of knowledge via rules realized by user-system transactions. Human-readable KBs are widely distributed via the world wide web and are typically accessed by search engines. Some search engines are used to mine data stored in KBs (Menasalvas et al. 2003). Knowledge-based systems incorporate artificial intelligence techniques to aid in a decision-making process (Leondes 2000, Palade et al. 2003).

KB-offered services vary widely. Yet, there are some common aspects, which are often seen as problems from the users' viewpoint. Complaints are linked with poor performance, as exemplified by long waiting times, connection failures, and so on. Thus, performance is often described as inadequate or unacceptable. For KB management, typical problems are slow system response times, low productivity, and sometimes system saturation.

The purpose of this paper is the of study responsiveness when information is requested from distant KB hosts. The method consists of three elements: (i) an operational view of user-system communication, (ii) the introduction of a new performance measure called *"responsiveness to service requests"*, and (iii) the performance analysis of workflow. Responsiveness is obtained in closed form and its evaluation requires a few input data. Critical points indicating congestion are also obtained analytically. Finally, an illustrative example serves as a framework for discussing performance improvement.

**Definitions**

Let us consider Figure 1. The link shown contains a finite number of servers, denoted by $S_1, S_2, S_3, ... , S_K$, the last server being the KB host. Users occupy $M$ workstations from which they generate their requests. At any time, a portion of $M$ users, denoted by $N$, will be active while the rest $M$-$N$ will be in a thinking state. Requests come form the active users.

The routing of requests follows the direction of the arrows. The communication system is assumed capable of servicing $N$ requests with $N$ constant during some period of usage. This assumption is realistic when the above is a period of heavy demand; then a backlog of requests is formed in the users' area. When a request completes service, the next in-line request is admitted via the input/output port: this action keeps $N$ constant. Periods of heavy system usage are especially interesting from a management perspective.

From the above it follows that the traffic at the I/O port should always be regulated on-line, and this is true with all protocols in use today. Such active regulation keeps $N$ practically constant at a number called the *window size*. Later we explore a range of values for $N$ and then find the maximum allowable value, when the system becomes congested and thus liable to performance saturation.

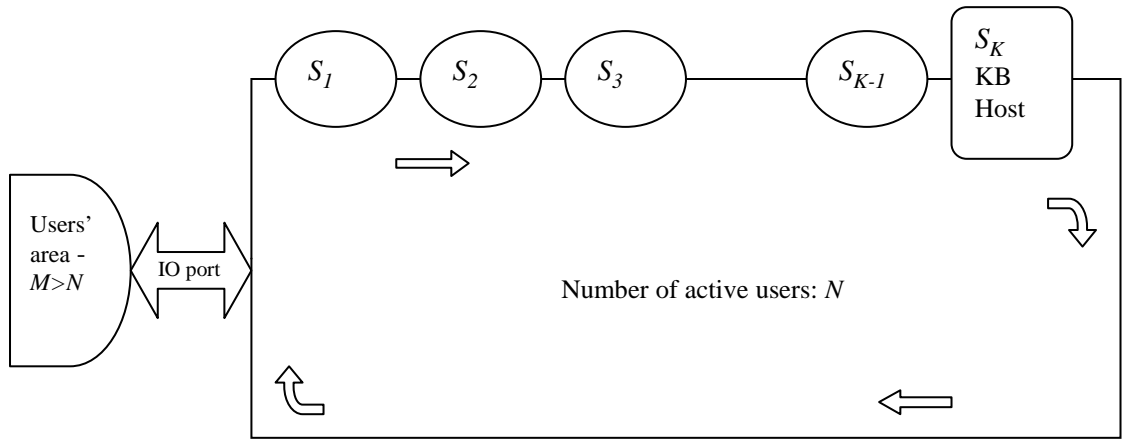

**Figure 1: Cyclic queueing model of a communication system with a KB host.**

With reference to Figure 1, we define the following performance metrics:

(i) *Total service ability of system, $\sigma(s)$*: the sum of the service times ($s_i$) of its servers.
(ii) *Elapsed time, $E(N)$*: time required for service completion including queueing delays.
(iii) *Think time, $T(u)$*: user time from instant of service completion to issue of next request.

From these metrics we may now introduce the following composite performance measure:
*Responsiveness to service requests, $R(s)$* = {total service ability of system}/
{above quantity augmented by the corresponding mean elapsed time}, i.e.:

$$R(s) = \sigma(s)/\{\sigma(s) + E(N)\} \qquad (1)$$

## Operational analysis

A performance model has a central role in system evaluation: it gives estimates about the system's operation and it can also provide management with performance predictions under alternative choices. When a capacity planning exercise or system redesign is due, such predictions can be of significant value.

*Modelling alternatives*

Several forms of performance models have been developed over the years. Earlier models were purely empirical, based on experimental methods from statistics such as regression. Simulation has also been a favourite technique amongst many performance analysts. Analytical models based on *stochastic analysis* (queueing theory) are a better alternative. They became widely acceptable, and remain so, because of the following reasons: (a) they are easier to construct and validate, (b) they can be realized using standard programming languages, and (c) results are given in closed-form expressions with clearer interpretations.

An interesting departure from classical stochastic theory is *operational analysis*. This method retains the basic form of its predecessor, including the advantages listed above. However, it is based upon a group of so-called operational rather than probabilistic laws. For this work, we have opted for a simple form of operational analysis, which allows for the study of user-system interaction with minimal computational effort; it is also useful for the examination of system transient behaviour during peak use.

The literature on queueing systems is quite extensive. We mention briefly the book by Bose (2002) which contains a thourough coverage of open and closed networks. The edited volume by Haring et al. (2000) contains a large collection of papers on the performance evaluation of computer databases and communication networks.

*Evaluation of responsiveness*

The total service ability of the system is by definition $\sigma(s) = (s_1+s_2+\cdots+s_K)$: this quantity can be evaluated from known server specifications. Therefore, the main task here is to estimate $E(N)$ in eq.(1). An exact estimation is possible using operational analysis; however, the procedure is iterative with each value of $E(N)$ depending on previous values. An acceptable approximation in closed-form is preferable as this would give a direct estimate of $E(N)$ for any value of $N$. Such an approximation could be obtained by taking into account the systems' maximum throughput as follows.

Let $\rho_i$ be the utilization of server $S_i$ and $\gamma_i$ its mean throughput. In operational analysis, this utilization can be expressed as the product of the throughput and the corresponding service time, i.e. $\rho_i = \gamma_i \cdot s_i$. As the system becomes loaded with larger values of $N$, the queues of requests in front of the servers will become longer. Eventually, there will be some instant when the slowest server will have to complete work at its maximum: at that instant, its utilization will have reached 100%. This saturated server becomes a limiting factor or bottleneck of the system. Let us denote by $S_{max}$ the bottleneck server, by $\rho_{max}$ its utilization and by $\gamma_{max}$ its throughput. $S_{max}$ is the slowest server because it has the largest mean service time, which we define by $s_{max}= \max \{ s_1, s_2, ... , s_K\}$. Then, as in the case of the mean values: $\rho_{max}= \gamma_{max} \cdot s_{max} = 1$. Therefore, $\gamma_{max}= 1/s_{max}$.

Let $\gamma(N)$ be the mean system throughput, i.e. the mean rate (in seconds) at which requests leave server $S_K$ with $N$ requests present. Since $\gamma_{max}$ is the maximum throughput anywhere in the system, it follows that the maximum value of $\gamma(N)$, denoted by $\gamma^*$, could be closely approximated by $\gamma_{max}$, i.e. $\gamma^* \approx \gamma_{max} = 1/s_{max}$. A well-known property of operational analysis states that in a closed queueing system the number of "customers" being served is equal to the product of the system's throughput and the waiting time (with all quantities as means). This expression is an adaptation of Little's law from classical stochastic analysis.

In the present notation, the mean number of "customers" are the user requests $N$, the mean system throughput is $\gamma(N)$, and the mean waiting time is $E(N)$. Little's law will also apply when $\gamma(N)$ has reached its maximum, in which case: $N = \gamma^* \cdot E(N)$. Substituting $\gamma^*$ as $\gamma^* \approx 1/s_{max}$ and then solving for $E(N)$ the above becomes the following approximation: $E(N) \approx N \cdot s_{max}$. Let us return to eq.(1). The quantity $\sigma(s)$ was evaluated before, and $E(N)$ was approximated above by $N \cdot s_{max}$. Eq.(1) now takes the following form:

$$R(s) \approx \sigma(s) / \{\sigma(s) + N \cdot s_{max}\} \qquad (2)$$

Note that the equal sign in eq.(1) has become a near equal sign due to the approximation introduced previously. When the system is empty, i.e. $N = 0$, $R(s) \approx 1$: this is an ideal case (i.e. 100% responsiveness), which cannot be expected in an operational system. When activity increases from $N = 1$ onwards, $R(s)$ decreases proportionately as in eq.(2). In the limiting case when $N \rightarrow \infty$, which practically means that $N$ has exceeded some critical point (to be determined), it follows that $R(s) \rightarrow 0$. This is be seen by the users as inability to communicate with the distant KB host.

### *Congestion and saturation*

When $N=1$ there is no contention; therefore, $E(1)$ reduces to the sum $\sigma(s)$. Little's law implies that $\gamma(1) \cdot E(1) = 1$ or $\gamma(1) = 1/\sigma(s)$. This is the minimum value of throughput. The corresponding maximum value was found to be approximately $1/s_{max}$. Therefore, the mean throughput is constrained as follows: $1/\sigma(s) \leq \gamma(N) \leq 1/s_{max}$. Thus, $\sigma(s) \cdot \gamma(N) \leq \sigma(s)/s_{max}$. As already stated, $\sigma(s) \cdot \gamma(N)$ is the mean number of "customers" $N$; therefore, $N \leq \sigma(s)/s_{max}$. Then, as $N$ increases, which means that user transactions ($n$) also follow, there will be some point, denoted by $N^*$, for which this inequality will eventually become an equality. Then, for such an $N = N^*$:

$$N^* \approx [\sigma(s)/s_{max}] \cdot n \qquad (3)$$

The near-equal sign in eq.(3) comes from the approximation $\gamma^* \approx 1/s_{max}$. The square brackets denote augmentation of the fraction to the next integer value. $N^*$ is called the critical point of the system: admission of requests $N$ when $N > N^*$ will increase congestion and then lead to saturation.

As previously discussed, requests for service are generated in the users' area. The mean throughput generated from $M-N$ users with an average think time $T(u)$ will be $(M-N)/T(u)$. Then, the ratio $\lambda_{in} = (M-N)/T(u) \cdot n$, will give the mean input rate into the system. When $N$ reaches its critical point $N^*$, $M$ will also reach a corresponding point $M^*$ in the users' area. Then, the system will have reached its maximum throughput $\gamma^* \approx 1/s_{max}$. Therefore, the corresponding maximum input rate $\lambda^*_{in}$ will be: $\lambda^*_{in} = (M^*-N^*)/T(u) \cdot n \approx 1/s_{max}$. Solution for $M^*$ implies that:

$$M^* \approx N^* + [T(u) \cdot n / s_{max}] \qquad (4)$$

The square brackets denote as before augmentation of the fraction to the next integer value. $M^*$ could be interpreted as follows. For an internal (i.e. inside the system) critical point $N^*$ there is a corresponding external critical point $M^*$, which shows the number of users when saturation appears. This will be evident at the slowest server. Therefore, the pair $(N^*, M^*)$ is an index of the system's ability to accommodate effectively its population of users.

**Limits of responsiveness**

Information retrieval involving $K$ servers is a good example of user-host communication. A link between a group of local users and their host may then be formed as suggested by Figure 1. Servers $S_1, S_2, \ldots, S_{K-1}$ are used for the relay of user queries and $S_K$ houses the distant KB. Service at $S_K$ includes the scheduling of user queries, searching in system files, and internal communications. For illustration we assume that $K = 15$. Their mean service times $s_i$ (in seconds) are chosen in the range $(0, 1)$, which reflects a wide choice of server processing speeds. Table 1 below shows the relevant data.

The slowest server is $S_7$ with $s_{max} = s_7 = 0.965$ seconds. The system's total service ability is $\sigma(s) = 10.140$ seconds. Application of eq.(2) for several values of $N$ has given the results for $R(s)$ shown in Table 2. Assuming a mean of $n = 10$ number of transactions per session, application of eq.(3) gives $N^* = 110$. Also, assuming that the average user needs $T(u) = 15$ seconds of think time, application of eq.(4) gives $M^* = 270$. Then, the pair (110, 270) shows the critical points of the system. The KB host should be allowed to serve up to 270 users of which 110 should be active at any time.

From Table 2 we note that for $N = 11$, responsiveness stands at $R(s) = 48.9\%$, i.e. at less than half of its ideal value. If more requests are allowed into the system, $R(s)$ will continue to decline moderately, as shown in Table 2 for $N$ from 12 to 15. Going further up to the critical point $N^* = 110$, we note that $R(s)$ declines sharply and now stands at $R^*(s) = 8.75\%$. In this limiting case, the maximum number of users are all active thus bringing the whole system to saturation.

| Service | Value | | | | |
|---|---|---|---|---|---|
| $s_1 ..... s_5$ | 0.546 | 0.467 | 0.847 | 0.325 | 0.645 |
| $s_6 ..... s_{10}$ | 0.835 | 0.965 | 0.628 | 0.617 | 0.564 |
| $s_{11} .... s_{15}$ | 0.873 | 0.674 | 0.694 | 0.726 | 0.734 |
| $s_{max} = s_7 = 0.965$ | | $\sigma(s) = 10.140$ | | | |

**Table 1: Service times in seconds for the *K* servers.**

| $N$ (users) | | $R(s)$ (%) | | | | |
|---|---|---|---|---|---|---|
| 1 ..... 5 | : | 91.3 | 84.0 | 77.8 | 72.4 | 67.8 |
| 6 ..... 10 | : | 63.7 | 60.0 | 56.8 | 53.9 | 51.3 |
| 11 ..... 15 | : | 48.9 | 46.7 | 44.7 | 42.9 | 41.2 |
| $N^* = 110$ | $M^* = 270$ | $R^*(s) = 8.75$ | | | | |

**Table 2: Responsiveness to service requests as *N* increases.**

### *Workflow balance*

In the example, the slowest server $(S_7)$ is placed about halfway through $S_1$ and the KB host. If one attempts to eliminate the bottleneck at $S_7$ by replacing this server by a faster one, then a new bottleneck could appear elsewhere. This will probably be at $S_{11}$ as this server is the second slowest after $S_7$ (see Table 1). In fact, any cyclic system that is not properly regulated may contain several bottlenecks. Therefore, a global strategy will be needed to keep workflow unrestricted on an end-to-end basis. This may be seen as a problem of assigning input flows to servers in proportion to their service ability.

This is a complex problem in its general form. Solution could be obtained by optimization; but, even so, the solution will have to be adjusted periodically as the workflow changes dynamically following user behaviour. We will not address this complex problem here. Readers may see, for instance, Giokas and Pentzaropoulos (2000) for solutions based on multi-criteria methods.

### *KB host as the slowest server*

From Table 1, the following sequence is evident: $s_7 > s_{11} > s_3 > s_6 > s_{15}$. Therefore, $S_7$ is the slowest server, and the system's throughput is dominated by it. When $N = N^* = 110$, throughput is at its maximum, i.e.: $\gamma^* \approx 1/s_{max} = 1/s_7 = 1/0.965 = 1.036$. Let $\rho_7$ be the utilization (fraction of busy time) of $S_7$. From operational analysis, this quantity is the product of the throughput and corresponding mean service time; therefore, $\rho_7 = \gamma^* \cdot s_7 = 1$. This indicates that $S_7$ is saturated, which was expected, because $N$ above was assumed to have reached its critical point. By comparison, $\rho_K = \rho_{15} = \gamma^* \cdot s_{15} = 1.036*0.734 = 0.769$. Therefore, $\rho_K < 1$, which indicates that the host $S_K$ is in steady-state.

Let us now assume that $S_K$ becomes the slowest server. This is done by interchanging the values of $s_7$ and $s_{15}$ in Table 1. Then, $\rho_K$ becomes 1, which confirms that the host is now the saturated server. Our goal here is to arrive at a new utilization factor for $S_K$, say $\rho_K$(new), which would bring $S_K$ back to its steady-state. To achieved that, we will need to study the host in (partial) isolation from the rest of servers. The isolation is achieved by a technique known as *decomposition*. This technique is best known from Econometrics, where it has been applied in the analysis of large systems with many variables. Later, it was applied in the study of other complex systems including computer and communication systems.

For a closed queueing network with $K$ servers that models a large system with a KB host, the principle of decomposition may be stated as follows. Consider one server, e.g. $S_K$, and then replace the network by a two-server system only consisting of $S_K$ and one composite, flow-quivalent server, say $S_e$. This server is *flow-equivalent* to $S_1, S_2, \ldots, S_{K-1}$ in the sense that the throughput of the $K-1$ system equals the arrival rate at $S_K$ with the same number $N$. Assume that the decomposition principle is applied to the network of Figure 1. Note that the throughput of $S_e$, representing the first $K-1$ servers, equals the arrival rate at $S_K$ (host). Calling the first quantity $\gamma_e(K-1)$ and the second one $\lambda_K$, we have: $\lambda_K = \gamma_e(K-1)$. The mean value of $\gamma_e(K-1)$ is not known. However, we may use the arguments of the previous section to estimate its maximum value, say $\gamma_e^*(K-1)$. This is done by observing that the slowest server in the $K-1$ system is $S_{11}$, with $s_{11} = 0.873$. Then, as in the case of the entire system studied previously, we may write: $\gamma_e^*(K-1) \approx 1/s_{max}(K-1) = 1/s_{11} = 1/0.873 = 1.145$.

From the above result $\gamma_e(K\text{-}1) \leq 1.145$. Let us assume for illustration that this unknown mean is about three-quarters of its maximum, i.e.: $\gamma_e(K\text{-}1) \approx (3/4)\cdot\gamma_e^*(K\text{-}1) \approx (3/4)*1.145 \approx 0.859$. Note that $s_K$ is now 0.965, because of the interchange between $s_7$ and $s_K$, which makes the host the slowest server. From the results obtained, the new utilization of $S_K$, say $\rho_K(\text{new})$, may be obtained as: $\rho_K(\text{new}) = \lambda_K \cdot s_K = \gamma_e(K\text{-}1)\cdot s_K = 0.859*0.965 = 0.829$. We easily check that $\rho_K(\text{new}) < 1$, which indicates that the host is brought back to its steady-state. The equation:

$$\lambda_K = \gamma_e(K\text{-}1) \qquad \textit{(flow-equivalent system)} \tag{5}$$

is a sufficient operational condition for achieving flow balance between the KB host and the $K$-1 servers.

## Concluding remarks

The purpose of this paper was the study of responsiveness in user-KB communications taking place via remote links. Analysis was carried out by a cyclic queueing model based on operational analysis. The performance measure introduced was named *"responsiveness to service requests"*. $R(s)$ was obtained in a simple, closed-form expression: its evaluation required only a few input data, and the calculations were direct, i.e. without any iterations. Critical points $(N^*, M^*)$, which indicate system congestion, were obtained analytically. The performance of the KB host was also studied by the application of decomposition and flow-equivalent aggregation. The equation "$\lambda_K = \gamma_e(K\text{-}1)$" is then a sufficient condition for achieving flow balance between the KB host and the $K$-1 servers.

Responsiveness was considered here as a measure of service quality: as such, it may also be seen as an index of user satisfaction. When $R(s)$ is within normal limits, users should appreciate the benefits from using the services offered by their host. When $R(s)$ declines, so does the picture of the communication system, as seen by the its users. Limits on the number of admissible requests and, whenever possible, on the number of connected users were considered necessary in order to avoid system congestion and possible saturation. Equating the flow out of $S_e$ with the flow into $S_K$ ensures that no one dominates the other.

The above results suggest the following rule of good practice for managing the KB host ($S_K$) effectively. *If the host is the slowest server, its input rate should be lowered until it matches the throughput of a composite server ($S_e$), which is flow-equivalent to the K-1 request-relaying servers of the system.*

## Acknowledgements

The present work has been partially financed by grant no. 70/4/4733, awarded by the Research Committee of the University of Athens, Greece.

___________________________________________________________________________